\documentclass[useAMS,usenatbib]{mn2e}
\usepackage{graphicx}
\usepackage{amsmath}
\usepackage{amssymb}
\newcommand\aj{{AJ}}%
\newcommand\araa{{ARA\&A}}%
\newcommand\apj{{ApJ}}%
\newcommand\apjl{{ApJ}}%
\newcommand\apjs{{ApJS}}%
\newcommand\aap{{A\&A}}%
\newcommand\mnras{{MNRAS}}%
\newcommand\pasp{{PASP}}%
\def\nad{\mbox{Na\,{I}\,D}}
\def\nadone{\mbox{Na\,{I}\,D$_1$}}

\newcommand\ion[2]{#1$\;${\small\rmfamily\@Roman{#2}}\relax}%
\title [Low Resolution \nad\ is a Bad Proxy for Extinction]{Low-Resolution Sodium D Absorption is a Bad Proxy for
  Extinction}

\author[Poznanski et al.]
{Dovi Poznanski$^{1,2,3}$\thanks{dovi@berkeley.edu},
Mohan Ganeshalingam$^{2}$,
Jeffrey M. Silverman$^{2}$,\newauthor
and Alexei V. Filippenko$^{2}$\\
\\
$^{1}$Lawrence Berkeley National Laboratory, 1 Cyclotron
  Road, Berkeley, CA 94720.\\
$^{2}$Department of Astronomy, University of California,
  Berkeley, CA 94720-3411.\\
$^{3}$Einstein Fellow.\\}

\begin{document}
	
\maketitle

\label{firstpage}
\begin{abstract}

Dust extinction is generally the least tractable systematic uncertainty in astronomy, and particularly in supernova science. Often in the past, studies have used the equivalent width of \nad\ absorption measured from low-resolution spectra as proxies for extinction, based on tentative correlations that were drawn from limited data sets. We show here, based on 443 low-resolution spectra of 172 Type Ia supernovae for which we have measured the dust extinction as well as the equivalent width of \nad, that the two barely correlate. We briefly examine the causes for this large scatter that effectively prevents one from inferring extinction using this method.

\vspace{1cm}
\end{abstract}

\vspace{1cm}
\section{Introduction}

The ubiquity of dust in galactic environments, most notably in star forming regions where many supernovae (SNe) explode, makes extinction correction one of the most pervasive and probably the least tractable systematic uncertainty in the study of SNe. When an object's intrinsic colors are unknown, as occurs frequently with rare and new classes, one is bound to look for indirect measurements of the amount of extinction and reddening that need to be taken into account.

The absorption doublet of sodium \nad, at 5890 and 5896\,\AA, is a well-known tracer of gas, and its strength is generally expected to indicate the amount of dust along the line of sight. \citet{ferlet85}, for example, show that the column densities of \nad\ correlate with those of hydrogen in the diffuse interstellar medium. However, line saturation often makes density estimations arduous (e.g., \citealt{mugglestone65}). Variations in dust-to-gas ratios in different galaxy types (see, for example, \citealt{issa90,lisenfeld98}) make the jump from sodium to dust somewhat more uncertain. Depletion of metals on dust grains (e.g., \citealt{savage79}) further complicates the picture. It would therefore be quite surprising to find a tight correlation between the strength of sodium absorption and dust extinction.

\citet{richmond94} have shown, using 57 high-resolution stellar spectra compiled from the literature, that the equivalent width (hereafter EW) of the individual components of \nad\ does indeed correlate with the color excesses measured for these stars, with a noticeable scatter. \citet{munari97} add a body of 32 stars observed with a somewhat lower resolution ($R \approx$ 16,500 vs. $R \approx$ 60,000--600,000); they constrain the nonlinearity of the relation, extensively discuss its pitfalls (such as when there are multiple absorption components, or when the extinction is higher than $E(B-V) \approx 0.4$ mag), and find that the precision is limited to about $\delta E(B-V) \approx 0.05$--0.15\,mag. They also show how at $\mathrm{EW}>0.5$\,\AA\ the \nadone\ line saturates and the relation flattens.

Type Ia SNe --- understood to be the result of the thermonuclear disruption of a white dwarf (see, for example, \citealt{hillebrandt00} for a review) --- are very effective distance indicators through ``standardization'' of their light curves; they were used to discover the accelerating expansion of the Universe \citep{riess98,perlmutter99} and to accurately measure the Hubble constant \citep[][and references therein]{riess11}. Since their luminosity is found to correlate well with their light-curve shape \citep{phillips93}, various methods fit those light curves to templates in order to extract a distance (e.g., \citealt{riess99,guy05,conley08}). In the process, dust extinction --- effectively a nuisance parameter --- is treated in various ways, either fit separately or as part of a more generic color dispersion term.

\citet{barbon90} and \citet{turatto03} used this property of SNe~Ia to compare the EW of \nad\ measured in low- to medium-resolution spectra to the extinction they derived from light-curve fitting. This would extend the method to the significantly more common occurrence, when one does not have high-resolution spectra and the doublet is blended. Using few spectra in each case (6 and $\sim$30, respectively), they derived tentative scaling relations which have been widely used in the literature over the last two decades. However, as we show below through a similar analysis of a much larger sample, these relations are of little predictive power. While a correlation indeed exists between extinction and the EW of \nad, the very large scatter makes it essentially useless when only low-resolution measurements are available\footnote{See also a similar discussion by N. Elias de la Rosa at \texttt{http://online.kitp.ucsb.edu/online/snovae07/eliasdelarosa/} and \citet{elias-rosa07}.}. 

\section{Data and Measurement}

For over a decade the Lick Observatory SN search (LOSS; \citealt{li00,filippenko01}) with the Katzman Automatic Imaging Telescope (KAIT) has been one the most successful systematic discovery engines of nearby SNe. KAIT has also conducted follow-up photometry of hundreds of SNe, along with the Nickel 1-m telescope at the Lick Observatory. Moreover, the LOSS team has obtained spectra of these SNe with multiple telescopes, but mostly with the Kast double spectrograph \citep{miller93} mounted on the Lick Observatory 3~m Shane reflector, the Low Resolution Imaging Spectrometer \citep[LRIS;][]{oke95} mounted on the 10~m Keck telescopes, and the Deep Imaging Multi-Object Spectrograph \citep[DEIMOS;][]{faber03} on the 10~m Keck\,II telescope. The photometric data are presented by \citet{ganesh10}, and the spectroscopy by Silverman et al. (in preparation). In addition, we have compiled photometry from the Cal\'{a}n/Tololo SN search \citep{hamuy96} and the Harvard CfA samples \citep{riess99,jha06,hicken09}.

\subsection{Extinction Estimation}

Ganeshalingam et al. (in preparation; see also \citealt{ganesh10}) fit every SN~Ia to templates using the multicolor light-curve shape fitter, MLCS2k2.v006 (hereafter MLCS; \citealt{riess96,jha07}), providing an estimate of the amount of extinction suffered by the SN in its host galaxy. Milky Way reddening is removed by using the dust maps of \citet{schlegel98}. Two independent fits are performed, using a galactic line of sight prior extinction law with the Milky Way average of $R_V=3.1$ or a steeper $R_V=1.7$ \citep[e.g.,][and references therein]{wang09a,hicken09a}. While the two sets give systematically different values for the amount of extinction $A_V$, the results are highly correlated and do not affect our conclusions below. For simplicity, we discuss only the values derived with $R_V=3.1$, which is consistent the value of $R_V=2.8 \pm 0.3$ found by \citet{chotard11}.

While there may be some debate as to the accuracy of the derived extinction values with MLCS due to the dependence on the prior dust law and to the degeneracy between intrinsic color variance and external reddening, the precision has been widely tested and vetted, far beyond our needs for this work.

\begin{figure*}
	\center
\includegraphics[width=.8\textwidth]{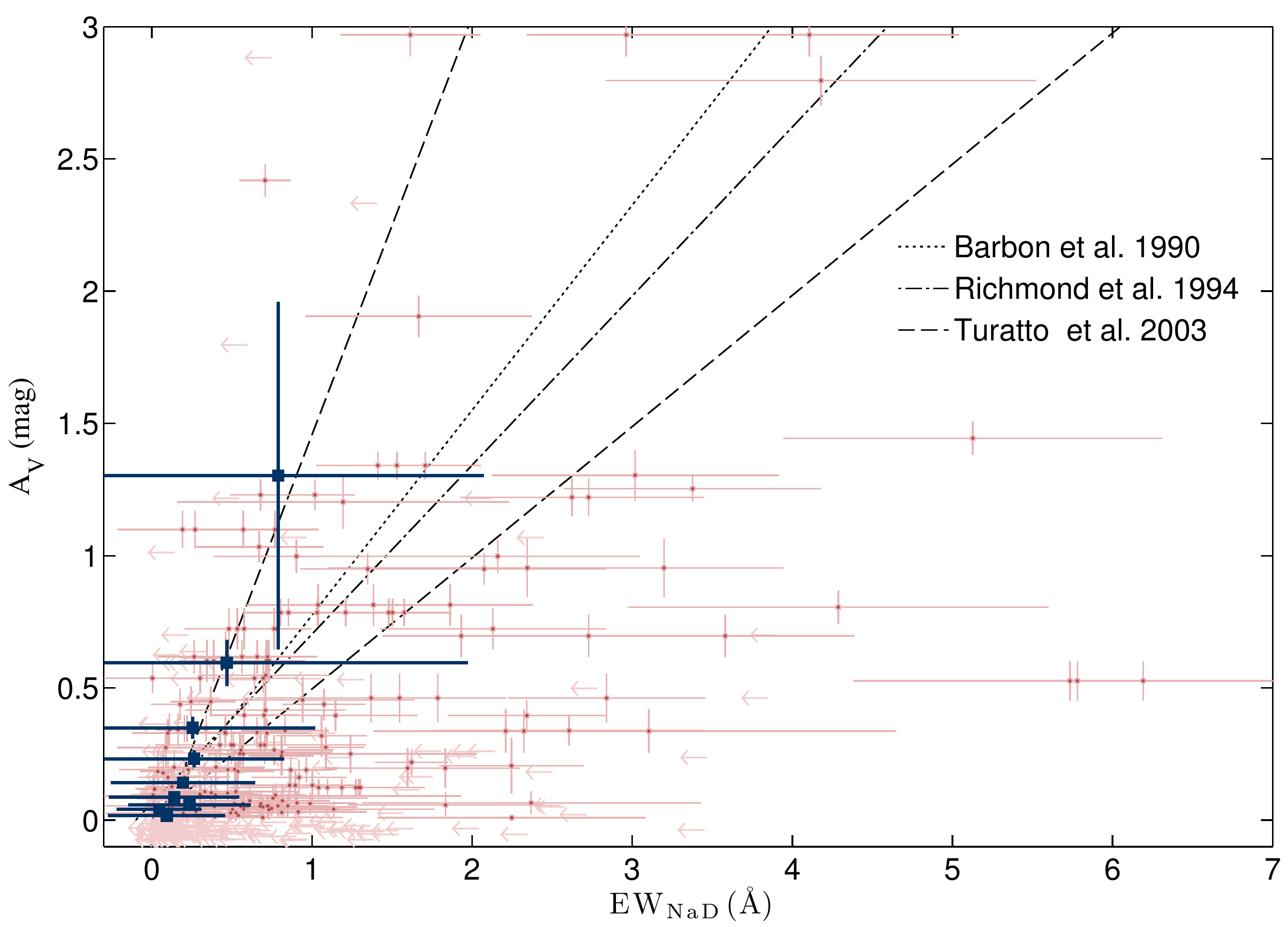}
\caption{In red, extinction along the line of
  sight, $A_V$, from template fitting to the light curves vs. the equivalent width of \nad\ as measured from low-resolution spectra of SNe~Ia. The large blue
  squares are weighted averages in bins of equal sample size. 
  Nondetections are marked by arrows located at the $1\sigma$
  limit. In addition we show the various correlations proposed by \citet{barbon90}, \citet{richmond94}, and \citet[][two different lines]{turatto03}. 
While there is indeed a correlation as seen in the mostly monotonic
  rise of the binned data, the scatter is far too great to make this
  spectral feature an effective predictor of dust
  extinction.\label{f:corr}}
\end{figure*}

\subsection{\nad\ Equivalent Width}

From our database of 1613 spectra of 743 SNe~Ia (Silverman et al., in preparation), we choose every spectrum of SNe having an MLCS estimate, resulting in 724 spectra of 239 individual SNe. In order to simplify the analysis we further cull the sample with the following criteria. (1) Objects that are found to be normal (i.e., non-peculiar) SNe~Ia based on matching with the SN identification code (SNID; \citealt{blondin07}). (2) Only spectra obtained with the three main spectrographs named above. (3) Redshift $z>0.005$, in order to avoid contamination of the EW measurement with observer-frame absorption from our own Galaxy. (4) MLCS goodness of fit of $\chi^2 < 2$, and estimated uncertainty in $A_V$ of $\delta A_V < 0.2$ mag. These cuts result in a sample of 443 spectra of 172 SNe. This sample is still more than an order of magnitude greater than in previous studies. Our results below are independent of the precise cuts applied (such as the redshift limit or minimal goodness of fit).

The procedure for deriving the EW of \nad\ for every spectrum is as follows. The spectrum is inspected visually, and if a line is noticeable its edges are manually marked. We perform numerical integration of the line over this marked range, dividing the result by the integration over the pseudo-continuum which is assumed to be linear. We divide the spectrum by a low-order polynomial fit to the rest-frame range of 5800--6000\,\AA\ and measure the noise of the spectrum, $N$, defined as the $5\sigma$ clipped standard deviation around the flattened spectrum. The uncertainty in the EW is assumed to be dominated by $N$. We calculate the uncertainty, $\delta$EW, as $\delta$EW $= N \times W/2$, where $W$ is the width (in \AA) of the line (the factor of 2 accounts for the approximately triangular shape of the feature). In cases where no line was detectable, we set EW $ = 0$, and for the uncertainty we use the average width for the line in our spectra (20\,\AA)\footnote{A table including all of our measurements is available at \texttt{http://astro.berkeley.edu/$\sim$dovi/Poznanski2011\_SN\_nad.dat}}.

\subsection{Variability}

For 85 of the SNe in our sample we have more than one spectrum, the majority with three or more spectra. For every such SN, we examine the stability of the EW of \nad\ by comparing the result for a given spectrum to the weighted mean for all spectra of that object. In only four cases (out of 254 spectra) do we find an EW that is inconsistent by more than $2\sigma$ with other spectra of the same SN. In two of those cases we find that this is due to a slight underestimation of the uncertainty. For the remaining two SNe, SN\,2006bq and SN\,2006cm, the variation in EW seems to arise from a varying amount of host-galaxy light spilling into the spectroscopic slit. This finding raises some concerns regarding the reliability of the EW of \nad\ as a good proxy for dust extinction, as measurements for the same SN can, on occasion, vary significantly.

\section{Correlation?}

In Figure \ref{f:corr} we show in red  the value of $A_V$ derived from MLCS compared to the EW of \nad. Nondetections are marked with arrows at the $1\sigma$ limit. We bin the EW measurements, with a variable bin size in order to have the same number of spectra per bin (about 50). We plot with large blue squares the weighted mean (in EW and $A_V$) for each bin. Here a shadow of the correlation arises, but the dispersion around it is too great to make it a useful tool. Our measurements are effectively consistent with all the tentative correlations suggested by \citet{barbon90}, \citet{richmond94}, and \citet{turatto03}, as shown in the figure. Our best fit correlation is $E(B-V) =  0.43\,\textrm{EW[\AA]} - 0.08$, with a systematic scatter of 0.3\,mag (1$\sigma$). A similar scatter is observed when plotting EW vs. MLCS $A_V$ values for $R_V=1.7$ or the SALT2 $c$ term \citep{guy07}. 

\begin{figure*}
	\center
\includegraphics[width=1\textwidth]{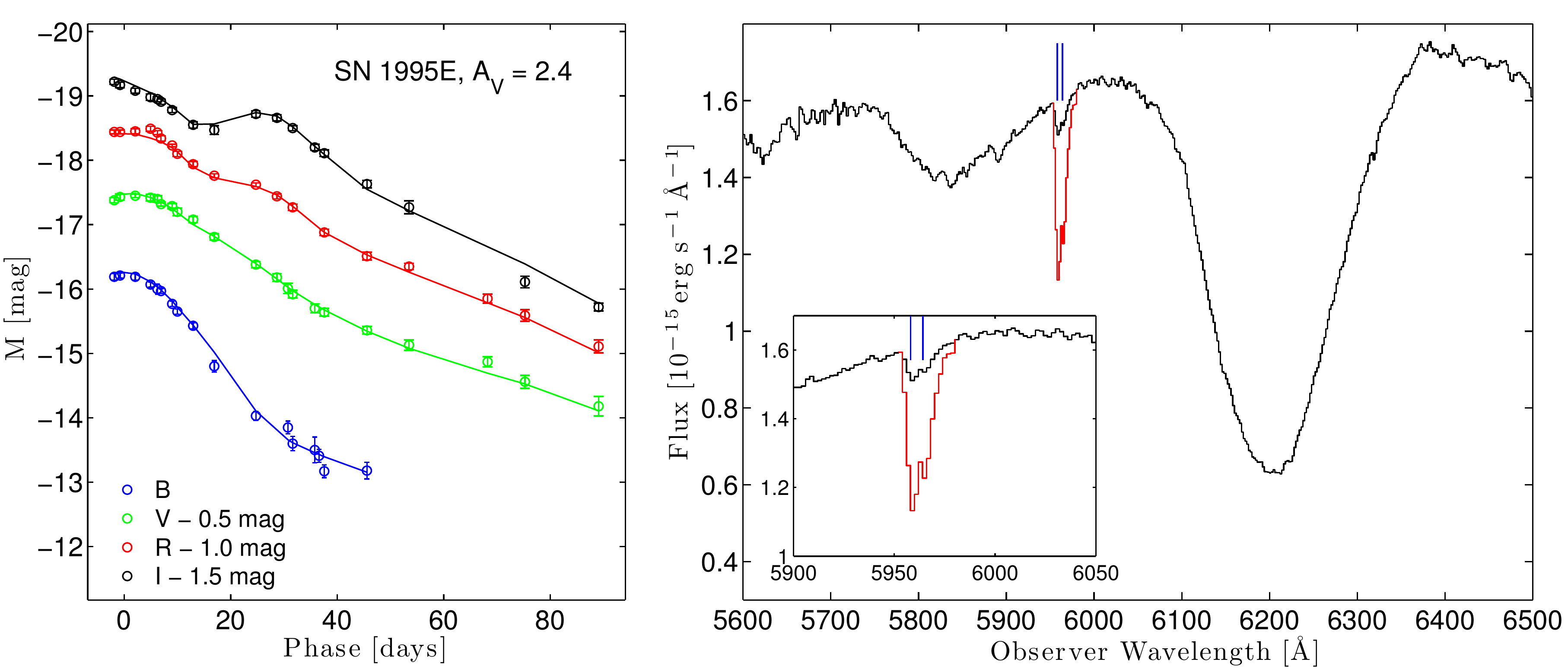}
\caption{Left: Multicolor light curves of SN\,1995E together with the
  best-fitting MLCS templates, from which a large extinction of
  $A_V=2.4$ mag is derived. Right: Low-resolution spectrum of
  SN\,1995E (inset shows the \nad\ feature in detail; blue lines mark
  the wavelengths of the doublet), indicating that the EW of \nad\ is
  rather small, about 0.7\,\AA. In red we show what the line should
  look like in order to have the EW predicted by the relation of
  \citet{richmond94}. \label{f:ex1}}
\end{figure*}

\begin{figure*}
	\center
\includegraphics[width=1\textwidth]{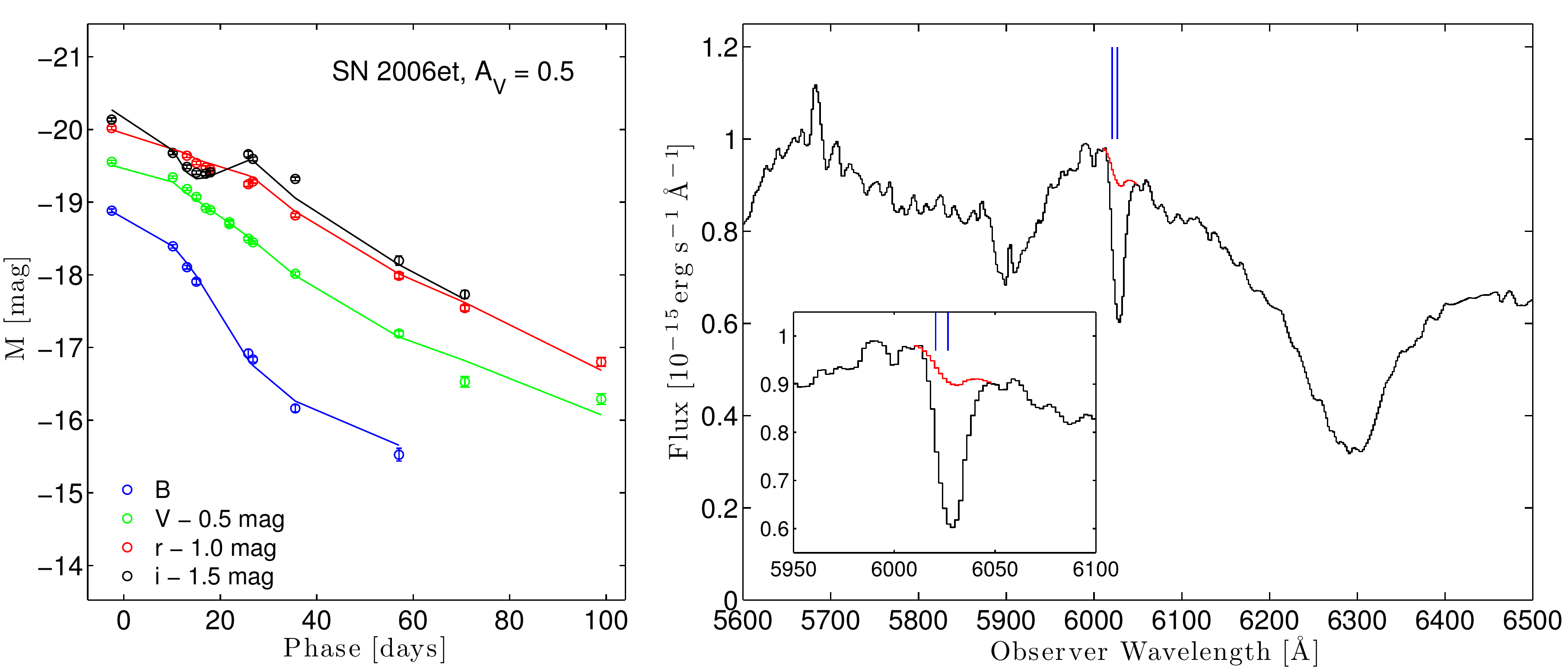}
\caption{Left: Multicolor light curve of SN\,2006et together with the
  best-fitting MLCS templates, from which a moderate extinction of
  $A_V=0.5$ mag is derived. Right: Low-resolution spectrum of
  SN\,2006et (inset shows the \nad\ feature in detail; blue lines mark
  the wavelengths of the doublet), indicating that the EW of \nad\ is
  extremely large, about 6\,\AA.  In red we show what the line should
  look like in order to have the EW predicted by the relation of
  \citet{richmond94}.\label{f:ex2}}
\end{figure*}

In Figures \ref{f:ex1} and \ref{f:ex2} we illustrate the two most extreme outliers in our sample, SNe 1995E and 2006et. Both are well fit by MLCS, with $A_V$ of 2.4 and 0.5 mag (respectively), but their \nad\ EW values are about 0.7\,\AA\ and 6\,\AA. While these are far from typical, Figure \ref{f:corr} shows that they are clearly not unique, and they demonstrate the risk associated with deriving extinctions from \nad\ EW measurements in low-resolution spectra.

\section{Conclusions}

Our finding is that the EW of \nad, as measured from low-resolution spectra, only very weakly tracks the amount of extinction suffered by a SN, at least as long as the extinction is in the range we probed, $A_V<3$ mag. Many effects conspire to minimize the amount of inference one could make. SN spectra are typically contaminated with host-galaxy light, and the continuum emission will dilute the line strength. This contamination is highly dependent on the slit width and position, as well as on observing conditions. 

In addition, low-resolution spectra cannot resolve the \nad\ doublet and therefore the different curves of growth of the two components are blended. Variations in gas-to-dust ratios in particular regions or galaxy types, multiple gas clouds at different velocities along the line of sight, and line saturation at large optical depths, all contribute to the theoretical and practical explanation for our results.

\section*{Acknowledgments}

We thank A.A. Miller, J.S. Bloom, and P.E. Nugent for useful comments on this manuscript. D.P. is supported by an Einstein Fellowship from NASA. The research of A.V.F.'s supernova group at UC Berkeley has been generously supported by the US National Science Foundation (NSF; most recently through grants AST--0607485 and AST--0908886), the TABASGO Foundation, US Department of Energy SciDAC grant DE-FC02-06ER41453, and US Department of Energy grant DE-FG02-08ER41563. KAIT and its ongoing operation were made possible by donations from Sun Microsystems, Inc., the Hewlett-Packard Company, AutoScope Corporation, Lick Observatory, the NSF, the University of California, the Sylvia \& Jim Katzman Foundation, the Richard and Rhoda Goldman Fund, and the TABASGO Foundation. Some of the data presented herein were obtained at the W.M. Keck Observatory, which is operated as a scientific partnership among the California Institute of Technology, the University of California, and NASA; the observatory was made possible by the generous financial support of the W.M. Keck Foundation. We thank the staffs of the Lick and Keck Observatories for their assistance with the observations.

\vspace{.5cm}
\bibliographystyle{mn2e} 
%\bibliography{myBIBTeX}

\end{document}